\documentclass[prd,twocolumn,tightenlines,showpacs,nofootinbib]{revtex4-1} 

\usepackage{amsfonts}
\usepackage{dcolumn}
\usepackage{bm}
\usepackage{amsmath}
\usepackage{nicefrac}
\usepackage{amsthm}
\usepackage{amssymb}
\usepackage{graphicx}
\usepackage{color}
\usepackage{url}
\usepackage{cancel}

\newcommand{\appropto}{\mathrel{\vcenter{
  \offinterlineskip\halign{\hfil$##$\cr
    \propto\cr\noalign{\kern2pt}\sim\cr\noalign{\kern-2pt}}}}}

\begin{document}

\title{Improved limits on interactions of low-mass spin-0 dark matter from atomic clock spectroscopy}

\date{\today}
\author{Y.~V.~Stadnik} 
\affiliation{School of Physics, University of New South Wales, Sydney 2052, Australia}
\author{V.~V.~Flambaum} 
\affiliation{School of Physics, University of New South Wales, Sydney 2052, Australia}

\begin{abstract}
Low-mass (sub-eV) spin-0 dark matter particles, which form a coherently oscillating classical field $\phi = \phi_0 \cos(m_\phi t)$, can induce oscillating variations in the fundamental constants through their interactions with the Standard Model sector. 
We calculate the effects of such possible interactions, which may include the linear interaction of $\phi$ with the Higgs boson, on atomic and molecular transitions. 
Using recent atomic clock spectroscopy measurements, we derive new limits on the linear interaction of $\phi$ with the Higgs boson, as well as its quadratic interactions with the photon and light quarks. 
For the linear interaction of $\phi$ with the Higgs boson, our derived limits improve on existing constraints by up to $2-3$ orders of magnitude. 
\end{abstract}


\maketitle

\textbf{Introduction.} --- 
Astrophysical observations indicate that dark matter (DM) is roughly 5 times more abundant (by energy content) than matter from the Standard Model (SM) \cite{Planck2015}, with a local cold (non-relativistic) DM energy density of $\rho_{\textrm{CDM}}^{\textrm{local}} \approx 0.4~\textrm{GeV/cm}^3$ determined from observations of stellar orbital velocities about our galactic centre \cite{Catena2010} and a present-day mean DM energy density of $\bar{\rho}_{\textrm{DM}} = 1.3 \times 10^{-6}~\textrm{GeV/cm}^3$ determined from measurements of the cosmic microwave background (CMB) radiation \cite{Planck2015}. 
Despite strong evidence for the existence of DM through its gravitational effects on SM matter, the identity and non-gravitational interactions of DM with the SM sector still remain unknown. 

Feebly-interacting, low-mass (sub-eV) spin-0 DM particles are a well-motivated candidate for DM. 
Arguably the most renowned particle that falls into this category is the axion, which is an odd-parity spin-0 particle that was originally proposed to resolve the strong CP problem of Quantum Chromodynamics (QCD), see, e.g., Ref.~\cite{Kim2010Axion_Review} for an overview. 
Apart from the axion, even-parity spin-0 particles, such as the dilaton (see, e.g., Refs.~\cite{Voloshin1987,Veneziano1988,Polyakov1994}) have also been conjuctured.
Low-mass spin-0 particles can be produced non-thermally in the early Universe via the `vacuum misalignment' mechanism \cite{Preskill1983cosmo,Sikivie1983cosmo,Dine1983cosmo}, and subsequently form a coherently oscillating classical field \cite{Footnote1}:~$\phi = \phi_0 \cos(\omega t)$, with an angular frequency of oscillation given by $\omega \simeq m_\phi c^2 / \hbar$, where $m_\phi$ is the mass of the spin-0 particle, $c$ is the speed of light and $\hbar$ is the reduced Planck constant. 
Unless explicitly stated otherwise, we adopt the units $\hbar = c = 1$ in the present work, with $1~\textrm{eV} = 2.4 \times 10^{14}~\textrm{Hz}$. 
Although typically produced with negligible kinetic energy, 
galactic spin-0 DM becomes virialised during galactic structure formation, which gives it the finite coherence time:~$\tau_{\textrm{coh}} \sim 2\pi / m_\phi v_{\textrm{vir}}^2 \sim 10^6 \cdot  2\pi / m_\phi $, i.e., $\Delta \omega/ \omega \sim 10^{-6}$.
This oscillating DM field bears the non-zero time-averaged energy density $\left< \rho_\phi \right> \simeq m_\phi^2 \phi_0^2 /2$ and satisfies the non-relativistic equation of state $\left< p_\phi \right> \ll \left< \rho_\phi \right>$, making it an ideal candidate for cold DM. 

If these spin-0 particles saturate the observed cold DM content, then their de Broglie wavelength must not exceed the DM halo size of the smallest dwarf galaxies ($R \sim 1$ kpc), which gives a lower bound on their mass:~$m_\phi \gtrsim 10^{-22}$ eV (though this limit is relaxed if the spin-0 particles make up only a sub-dominant fraction of the total cold DM). 
This simple estimate is in good agreement with more-detailed studies pertaining to the CMB and observed structure formation, see, e.g., Ref.~\cite{Marsh2015Review} for an overview. 
Ultra-low-mass spin-0 DM only behaves like perfect cold DM on length scales larger than its wavelength. 
On shorter length scales, gravitational collapse is prevented by quantum pressure \cite{Khlopov1985}. 
This phenomenon has non-trivial consequences for cosmology; in particular, it lies at the heart of ultra-low-mass DM models with particle masses in the range $10^{-24}~\textrm{eV} \lesssim m_\phi  \lesssim 10^{-20}$ eV, which have been proposed to resolve several long-standing ``small-scale crises'' of the cold DM model, see, e.g., Refs.~\cite{Hu2000,Marsh2014,Paredes2015}.

Apart from the mentioned effects in cosmological settings, low-mass spin-0 DM may also produce characteristic signatures in the laboratory. 
In particular, interactions of $\phi$ with the SM sector can induce variations in the fundamental constants \cite{Pospelov2010,Arvanitaki2015,Stadnik2015laser,Stadnik2015DM-VFCs}. 
Atomic and molecular spectrocopy offer powerful platforms to search for oscillating variations in the fundamental constants due to an oscillating DM field, and in the recent works \cite{Budker2015,Stadnik2015DM-VFCs,Bize2016}, new much-improved limits have already been obtained from atomic clock frequency comparison measurements.

In the present work, we calculate the effects of interactions of $\phi$ with the SM sector, which may include the linear interaction of spin-0 DM with the Higgs boson, on atomic and molecular transitions, and using recent atomic clock spectroscopy measurements, we derive new limits on several interactions of $\phi$ with the SM sector.
In the case of the linear interaction of $\phi$ with the Higgs boson, our results improve on existing constraints by up to $2-3$ orders of magnitude.

\textbf{Theory.} --- 
The field $\phi$ can couple to the SM fields in various ways, which include the following linear-in-$\phi$ interactions \cite{Footnote2}:
\begin{equation}
\label{lin_couplings_phi}
\mathcal{L}_{\textrm{int}}^{\textrm{lin}} = - \sum_{f}  \frac{ \phi }{\Lambda_f} m_f  \bar{f}f 
+ \frac{ \phi }{\Lambda_\gamma} \frac{F_{\mu \nu} F^{\mu \nu}}{4} \, ,
\end{equation}
where the first term represents the coupling of the spin-0 field to the SM fermion fields $f$, with $m_f$ the standard mass of the fermion and $\bar{f}=f^\dagger \gamma^0$, and the second term represents the coupling of the spin-0 field to the electromagnetic field tensor $F$, as well as the analogous quadratic-in-$\phi$ interactions: 
\begin{equation}
\label{quad_couplings_phi}
\mathcal{L}_{\textrm{int}}^{\textrm{quad}} = - \sum_{f}  \frac{ \phi^2 }{(\Lambda'_f)^2} m_f  \bar{f}f 
+ \frac{ \phi^2 }{(\Lambda'_\gamma)^2} \frac{F_{\mu \nu} F^{\mu \nu}}{4} \, .
\end{equation}
Comparing the terms in Eqs.~(\ref{lin_couplings_phi}) and (\ref{quad_couplings_phi}) with the relevant terms in the SM Lagrangian:
\begin{align}
\label{SM_Lagr+}
\mathcal{L}_{\textrm{SM}} \supset - \sum_{f}   m_f  \bar{f}f 
-  \frac{F_{\mu \nu} F^{\mu \nu}}{4} \, ,
\end{align}
we see that the linear-in-$\phi$ interactions in (\ref{lin_couplings_phi}) alter the fermion masses and the electromagnetic fine-structure constant $\alpha$ according to:
\begin{align}
\label{delta-SM_masses_lin}
m_f \to ~m_f \left(1 + \frac{\phi}{\Lambda_f} \right) , ~\alpha \to \frac{\alpha}{1 - \phi / \Lambda_\gamma } \simeq \alpha \left(1 + \frac{\phi}{\Lambda_\gamma} \right) \, , 
\end{align}
while the quadratic-in-$\phi$ interactions in (\ref{quad_couplings_phi}) alter the constants according to:
\begin{align}
\label{delta-SM_masses_quad}
&m_f \to ~m_f \left[1 + \frac{\phi^2}{(\Lambda'_f)^2} \right] \, , \\ \notag
&\alpha \to \frac{\alpha}{1 - \phi^2 / (\Lambda'_\gamma)^2 } \simeq \alpha \left[1 + \frac{\phi^2}{(\Lambda'_\gamma)^2} \right] \, , 
\end{align}

\begin{figure}[h!]
\begin{center}
\includegraphics[width=3.3cm]{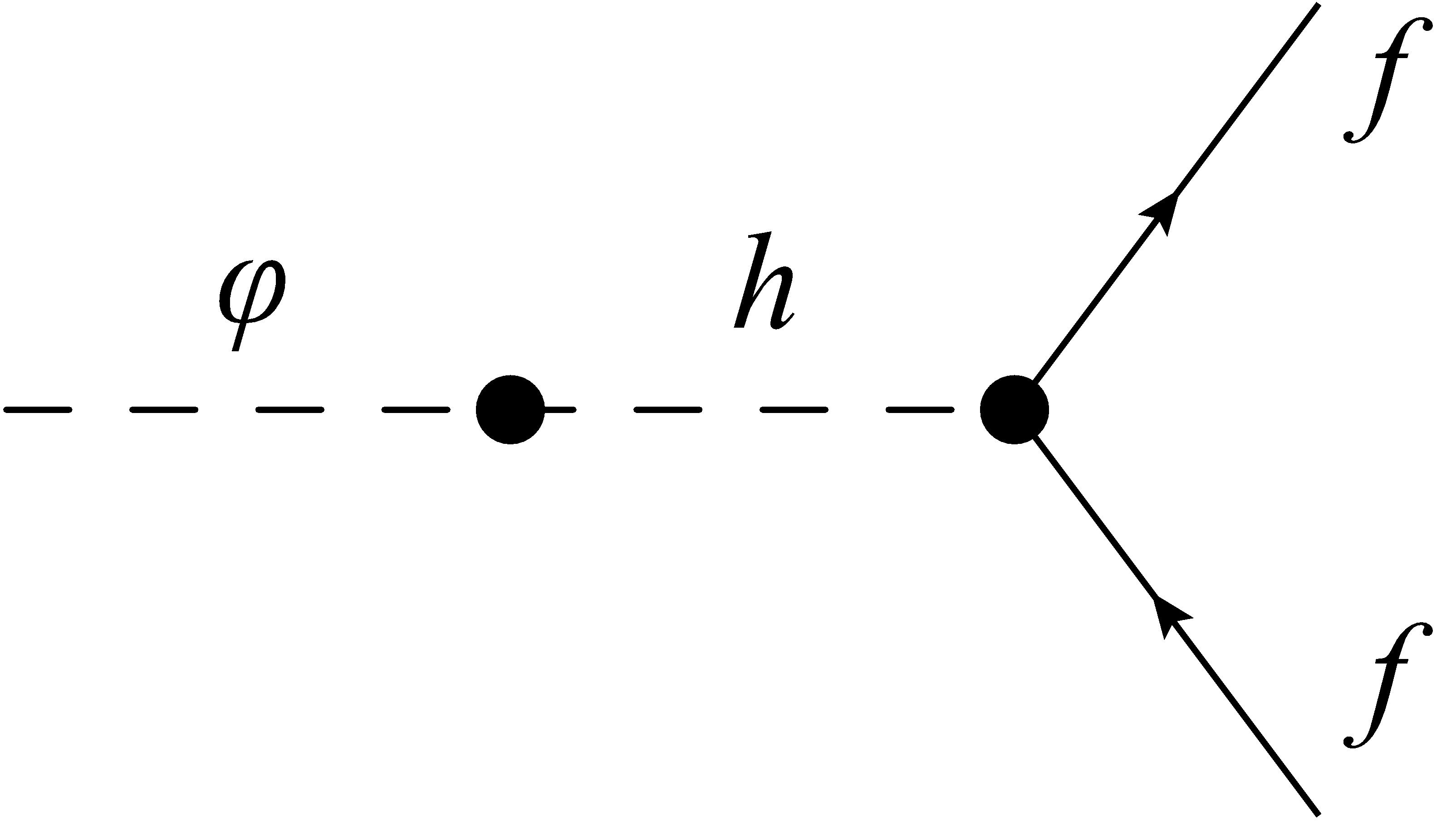} 
\includegraphics[width=4.8cm]{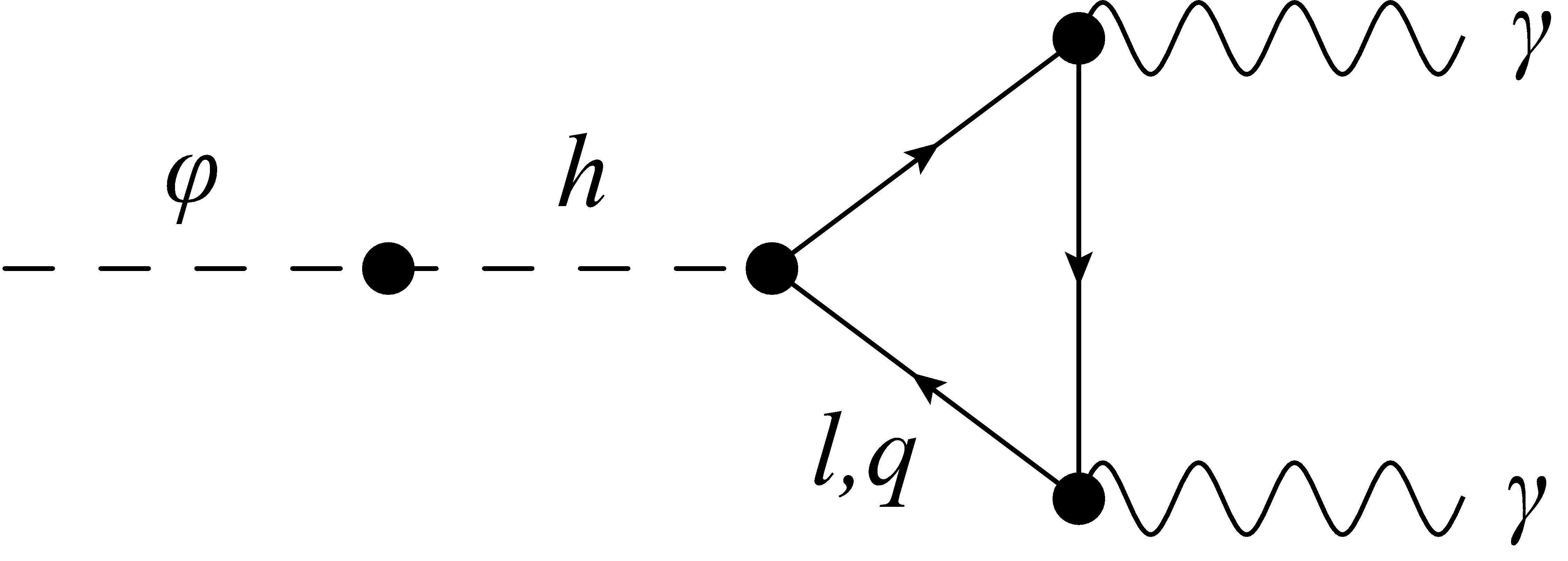} 
\caption{
Mixing of $\phi$ with the Higgs boson $h$ generates couplings of $\phi$ to the SM fermions (left) and radiatively to the SM electromagnetic field (right).} 
\label{fig:Higgs_couplings}
\end{center}
\end{figure}

The field $\phi$ may also couple to the Higgs field via the super-renormalisable interaction \cite{Footnote3}: 
\begin{equation}
\label{lin_couplings_phi_Higgs}
\mathcal{L}_{\textrm{int}}^{\textrm{Higgs}} = - A \phi H^\dagger H \, ,
\end{equation}
where $H$ is the Higgs doublet.  
To leading order in the interaction parameter $A$, the interaction (\ref{lin_couplings_phi_Higgs}) induces the following interactions of $\phi$ with the fermions and the electromagnetic field via mixing of $\phi$ with the physical Higgs field $h$ (see Fig.~\ref{fig:Higgs_couplings}) \cite{Pospelov2010}:
\begin{equation}
\label{lin_couplings_phi_Higgs-induced}
\mathcal{L}_{\textrm{int,eff}}^{\textrm{Higgs}} = \frac{A \left< h \right>}{m_h^2} \phi \left( \sum_f g_{hff} \bar{f}f 
+  \frac{g_{h\gamma\gamma}}{\left< h \right>}  F_{\mu \nu} F^{\mu \nu} \right) \, ,
\end{equation}
where $m_h = 125$ GeV is the mass of the Higgs boson, $g_{hff} = m_f / \left< h \right>$ for couplings of the Higgs to elementary fermions (leptons and quarks), $g_{hNN} = b m_N / \left< h \right>$ with $b \sim 0.2 - 0.5$ \cite{Shifman1978} for couplings of the Higgs to nucleons, and $g_{h\gamma\gamma} \approx \alpha / 8\pi$ for the radiative coupling of the Higgs to the electromagnetic field \cite{Pospelov1999-2002}.
Comparing the terms in Eq.~(\ref{lin_couplings_phi_Higgs-induced}) with the relevant terms in the SM Lagrangian (\ref{SM_Lagr+}), we see that the relevant fundamental constants are altered according to:
\begin{align}
\label{delta-SM_masses_lin_Higgs}
&m_f \to ~m_f \left[1 - \frac{A g_{hff} \left< h \right> \phi}{m_f m_h^2}  \right] \, , \notag \\
&\alpha \to \alpha \left[1 + \frac{4A g_{h\gamma\gamma}  \phi}{m_h^2} \right] \, .
\end{align} 

\textbf{Calculations.} --- 
The alterations in the fundamental constants due to the interactions of the oscillating field $\phi = \phi_0 \cos(m_\phi t)$ with the SM sector produce oscillating-in-time alterations in atomic and molecular transition frequencies, which depend on the various constants of nature. 
Most generally, the effect of variations of fundamental constants on the ratio of two clock frequencies, $\omega_1 / \omega_2$, can be written in the form: 
\begin{equation}
\label{general_clock_frequency_ratio}
\frac{\delta \left( \omega_1 / \omega_2 \right)}{ \omega_1 / \omega_2 } = \sum_{X} \left( K_{X,1} - K_{X,2} \right) \frac{\delta X}{X} \, ,
\end{equation}
where the sum runs over the dimensionless constants $X = \alpha, m_e / m_N, m_q / \Lambda_{\textrm{QCD}}$, which are the relevant combinations of physical constants when considering atomic and molecular transitions (see Eq.~(\ref{atomic_optical_frequency}) and the ensuing formulae of this section), and $K_X$ are the corresponding sensitivity coefficients. 
Here, $m_N = (m_p + m_n)/2$ is the averaged nucleon mass, $m_q = (m_u + m_d)/2$ is the averaged light quark mass, and $\Lambda_{\textrm{QCD}}$ 
is the QCD scale. 
We can present the oscillatory alterations in the ratio $\omega_1 / \omega_2$ resulting from Eqs.~(\ref{delta-SM_masses_lin}), (\ref{delta-SM_masses_quad}) and (\ref{delta-SM_masses_lin_Higgs}) due to an oscillating DM field $\phi = \phi_0 \cos(m_\phi t)$, for which $\rho_\phi \equiv \left< \rho_\phi \right> \simeq m_\phi^2 \phi_0^2 /2$, in the following respective forms \cite{Footnote4}:
\begin{equation}
\label{clock_frequency_ratio_lin}
\frac{\delta \left( \omega_1 / \omega_2 \right)}{ \omega_1 / \omega_2 } \approx \sum_{X=\alpha, m_e, m_q} \left( K_{X,1} - K_{X,2} \right) \frac{\sqrt{2 \rho_\phi}}{m_\phi \Lambda_X} \cos(m_\phi t) \, , 
\end{equation}
\begin{equation}
\label{clock_frequency_ratio_quad}
\frac{\delta \left( \omega_1 / \omega_2 \right)}{ \omega_1 / \omega_2 } \approx \sum_{X=\alpha, m_e, m_q} \left( K_{X,1} - K_{X,2} \right) \frac{\rho_\phi}{m_\phi^2 (\Lambda'_X)^2} \cos(2 m_\phi t) \, , 
\end{equation}
\begin{equation}
\label{clock_frequency_ratio_lin_Higgs}
\frac{\delta \left( \omega_1 / \omega_2 \right)}{ \omega_1 / \omega_2 } \approx \left( K_{H,1} - K_{H,2} \right) \frac{A \sqrt{2 \rho_\phi}}{m_\phi m_h^2} \cos(m_\phi t) \, ,
\end{equation}
where the sensitivity coefficient $K_H$ is defined as:
\begin{equation}
\label{coefficient_K_H}
K_H =\frac{\alpha }{2 \pi} K_\alpha - (1-b) K_{m_e} - 1.05 (1-b) K_{m_q} \, ,
\end{equation}
with $b \sim 0.2 - 0.5$.
In arriving at Eq.~(\ref{coefficient_K_H}), we have made use of the observation that the dominant physical-constant dependence of atomic/molecular transitions can only be through the combinations $\alpha$, $m_e/m_N$ and $\mu$, where $\mu$ is the dimensionless nuclear magnetic dipole moment in units of the nuclear magneton, with $\delta \ln(\mu) = K_{m_q} \delta \ln(m_q / \Lambda_{\textrm{QCD}})$, and have used the relations immediately succeeding Eq.~(\ref{lin_couplings_phi_Higgs-induced}), as well as the relation $\delta m_N / m_N \approx 0.05 ~ \delta m_q / m_q + 0.95 ~ \delta \Lambda_{\textrm{QCD}} / \Lambda_{\textrm{QCD}}$ (see, e.g., Refs.~\cite{Flambaum2004nuclear,Flambaum2005nuclear}). 
Formulae (\ref{clock_frequency_ratio_lin}) -- (\ref{coefficient_K_H}) provide a convenient parametrisation for extracting information on the various DM interaction parameters appearing in Eqs.~(\ref{lin_couplings_phi}), (\ref{quad_couplings_phi}) and (\ref{lin_couplings_phi_Higgs}), provided that one knows the relevant sensitivity coefficients.

An atomic optical transition may be presented in the following form:
\begin{equation}
\label{atomic_optical_frequency}
\omega_{\textrm{opt}} \propto \left( \frac{m_e e^4}{\hbar^3} \right) F_{\textrm{rel}}^{\textrm{opt}}(Z \alpha) \, ,
\end{equation}
where $F_{\textrm{rel}}^{\textrm{opt}}$ is a relativistic factor, which typically scales as $F_{\textrm{rel}}^{\textrm{opt}} \propto (Z \alpha)^2$, and can be calculated accurately via numerical many-body atomic calculations \cite{Dzuba1998A,Dzuba1998B,Dzuba2003,Angstmann2004,Dzuba2008,Flambaum2008Review}. 
We note that the non-relativistic atomic unit of frequency, $m_e e^4/\hbar^3$, appears in all atomic and molecular transition frequencies, and so cancels identically when considering the ratio $\omega_1 / \omega_2$.

An atomic hyperfine transition may be presented in the following form:
\begin{equation}
\label{atomic_hyperfine_frequency}
\omega_{\textrm{hf}} \propto \left( \frac{m_e e^4}{\hbar^3} \right) \left[ \alpha^2 F_{\textrm{rel}}^{\textrm{hf}}(Z \alpha) \right] \left(\frac{m_e}{m_N}\right) \mu \, ,
\end{equation}
where $F_{\textrm{rel}}^{\textrm{hf}}$ is the Casimir relativistic factor. 
For $s$ and $p$ states with $j=1/2$, the Casimir relativistic factor is given approximately by \cite{Prestage1995,Khriplovich1991}:
\begin{equation}
\label{relativistic_Casimir_factor}
F_{\textrm{rel}}^{\textrm{hf}} = \frac{3}{\gamma_{1/2} \left( 4\gamma_{1/2}^2 - 1 \right) } \, ,
\end{equation}
with $\gamma_{j} = \sqrt{(j+1/2)^2 - (Z \alpha )^2}$.
Variation with respect to $\alpha$ leads to the expression:
\begin{equation}
\label{relativistic_Casimir_factor_varn}
\frac{\delta F_{\textrm{rel}}^{\textrm{hf}}}{F_{\textrm{rel}}^{\textrm{hf}}} = K_{\textrm{rel}}^{\textrm{hf}} \frac{\delta \alpha}{\alpha} \, ,
\end{equation}
where $K_{\textrm{rel}}^{\textrm{hf}}$ is given by:
\begin{equation}
\label{relativistic_Casimir_factor_derivative}
K_{\textrm{rel}}^{\textrm{hf}} = \frac{\left(Z\alpha\right)^2 \left( 12 \gamma_{1/2}^2 - 1 \right)}{\gamma_{1/2}^2 \left( 4\gamma_{1/2}^2 - 1 \right) } \, .
\end{equation}
We note that more accurate numerical many-body calculations give slightly larger values of the coefficient $K_{\textrm{rel}}^{\textrm{hf}}$ than the analytical expression in (\ref{relativistic_Casimir_factor_derivative}) for moderately heavy atomic and ionic species \cite{Dzuba1998B,Tedesco2006nuclear}. 
The dependence of various nuclear magnetic dipole moments $\mu$ on the ratio $m_q / \Lambda_{\textrm{QCD}}$ has been calculated using a number of nuclear models, see, e.g., Refs.~\cite{Flambaum2004nuclear,Tedesco2006nuclear,Dinh2009nuclear}. 

In molecular transitions, there also exist rotational ($\Delta E_{\textrm{rot}} \propto m_e/m_N $) and vibrational ($\Delta E_{\textrm{vibr}} \propto \sqrt{m_e/m_N}$) degrees of freedom, in addition to the fine-structure and magnetic hyperfine contributions discussed above. 
In molecules, nearly-degenerate pairs of levels due to near cancellation of energy shifts of different nature arise quite often. In such cases, the relative sensitivity of the corresponding transition to variation in one or more of the fundamental constants may be significantly enhanced, $\left| K_X \right| \gg 1$, see, e.g., Refs.~\cite{Flambaum2006mol,Flambaum2007NH3,Flambaum2007mol,Ye2008mol,DeMille2008mol,Kozlov2012poly,Stadnik2013PbF,Pasteka2015}.

We present values of the sensitivity coefficient $K_H$, defined in Eqs.~(\ref{clock_frequency_ratio_lin_Higgs}) and (\ref{coefficient_K_H}), for a variety of atomic and molecular transitions, in Tables~\ref{tab:K_H_values_atom} and \ref{tab:K_H_values_mol}, respectively. 
From the tabulated values, it is evident that $|K_H| \ll 1$ for a typical atomic optical transition, while $|K_H| \sim 1$ for a typical atomic hyperfine transition. 
For molecular transitions, a large enhancement in the sensitivity coefficient is possible, $|K_H| \gg 1$.

\begin{table}[h!]
    \centering
    \caption{Calculated values of the sensitivity coefficient $K_H$, defined in Eqs.~(\ref{clock_frequency_ratio_lin_Higgs}) and (\ref{coefficient_K_H}), for selected atomic transitions.} 
\begin{tabular}{cccc}
\multicolumn{1}{c}{System} & Transition   & $K_H$ ($b=0.5$)	 & $K_H$ ($b=0.2$)    \\ 
\hline
$^{27}$Al$^{+}$ & $^{1}S_{0} \leftrightarrow$ $^{3}P_{0}$, optical		&	$9 \times 10^{-6}$	&	$9 \times 10^{-6}$	\\
$^{87}$Sr & $^{1}S_{0} \leftrightarrow$ $^{3}P_{0}$, optical	&	$7 \times 10^{-5}$	&	$7 \times 10^{-5}$	\\
$^{171}$Yb & $^{1}S_{0} \leftrightarrow$ $^{3}P_{0}$, optical		&	$4 \times 10^{-4}$	&	$4 \times 10^{-4}$	\\
$^{171}$Yb$^{+}$ &	$^{2}S_{1/2} \leftrightarrow$ $^{2}F_{7/2}$, optical	&	$- 7 \times 10^{-3}$	&	$-7 \times 10^{-3}$	\\
$^{171}$Yb$^{+}$ &	$^{2}S_{1/2} \leftrightarrow$ $^{2}D_{3/2}$, optical	&	$1 \times 10^{-3}$	&	$1 \times 10^{-3}$	\\
$^{199}$Hg & $^{1}S_{0} \leftrightarrow$ $^{3}P_{0}$, optical		&	$9 \times 10^{-4}$	&	$9 \times 10^{-4}$	\\
$^{199}$Hg$^{+}$ & $^{2}S_{1/2} \leftrightarrow$ $^{2}D_{5/2}$, optical		&	$-3 \times 10^{-3}$	&	$-3 \times 10^{-3}$	\\
$^{162}$Dy &  $4f^{10} 5d 6s \leftrightarrow 4f^{9} 5d^{2} 6s$	&	$1 \times 10^{4}$	&	$1 \times 10^{4}$	\\
$^{164}$Dy &  $4f^{10} 5d 6s \leftrightarrow 4f^{9} 5d^{2} 6s$	&	$-3 \times 10^{3}$	&	$-3 \times 10^{3}$	\\
$^{1}$H &	ground-state hyperfine	&	$-0.45$	&	$-0.72$	\\
$^{87}$Rb &	ground-state hyperfine	&	$-0.49$	&	$-0.78$	\\
$^{133}$Cs &	ground-state hyperfine	&	$-0.50$	&	$-0.80$	\\
\end{tabular}%
    \label{tab:K_H_values_atom}
\end{table}

\begin{table}[h!]
    \centering
    \caption{Calculated values of the sensitivity coefficient $K_H$, defined in Eqs.~(\ref{clock_frequency_ratio_lin_Higgs}) and (\ref{coefficient_K_H}), for various types of molecular transitions. The values presented are for the most sensitive known transitions and assume $b=0.2$.} 
\begin{tabular}{ccc}
\multicolumn{1}{c}{System}   & \multicolumn{1}{c}{Transition}   & $K_H$     \\ 
\hline
diatomic	&	rotational/hyperfine	&	$\sim 1$		\\
diatomic	&	$\Omega$-type doubling/hyperfine	 &	$\sim -3$		\\
diatomic	&	fine-structure/vibrational	&	$\sim 10^2$		\\  
linear polyatomic	&	various &	$\sim 10^3$		\\
diatomic cation	&	various &	$\sim - 10^6$		\\
NH$_{3}$	&	inversion	 &	$-3.6$		\\
\end{tabular}%
    \label{tab:K_H_values_mol}
\end{table}

\textbf{Results.} ---
Using the formulae of the previous section and the experimental data of Refs.~\cite{Budker2015,Bize2016}, we now derive new limits on the interaction parameters that appear in Eqs.~(\ref{quad_couplings_phi}) and (\ref{lin_couplings_phi_Higgs}). 
In doing so, we assume that the spin-0 DM field $\phi$ saturates the local galactic cold DM content ($\rho_{\textrm{CDM}}^{\textrm{local}} \approx 0.4~\textrm{GeV/cm}^3$ \cite{Catena2010}).

\begin{figure}[h!]
\begin{center}
\includegraphics[width=8.5cm]{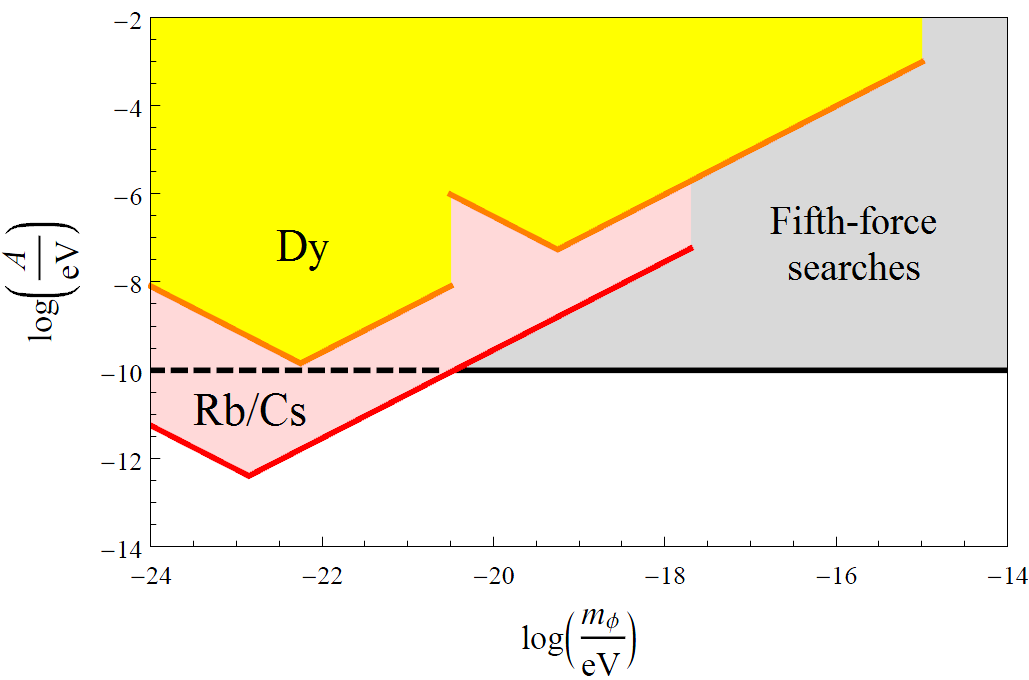} 
\includegraphics[width=8.5cm]{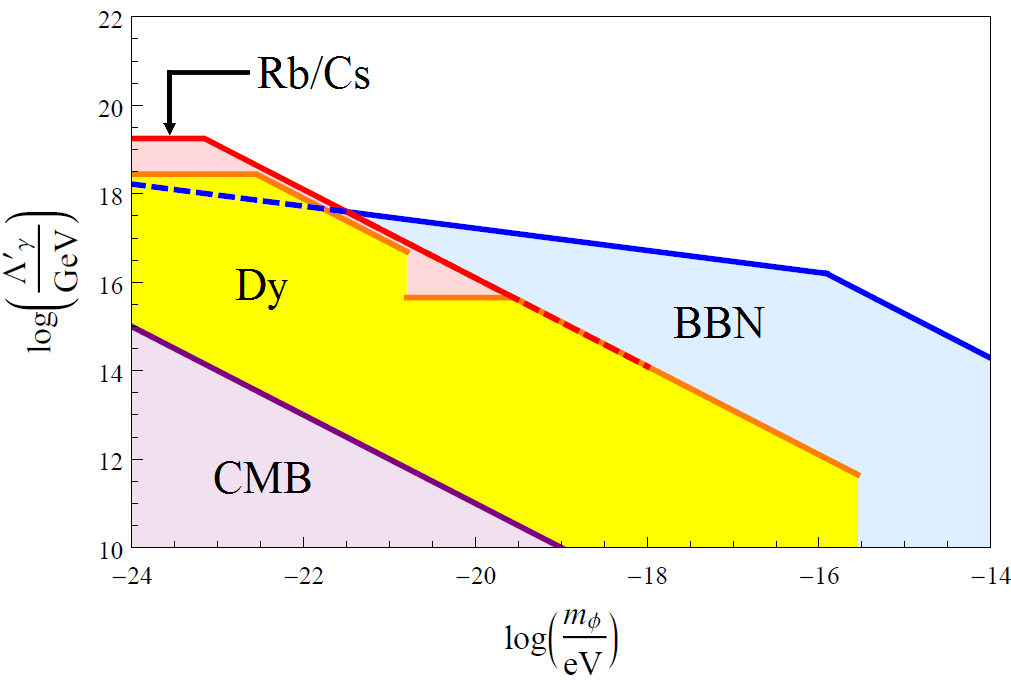} 
\includegraphics[width=8.5cm]{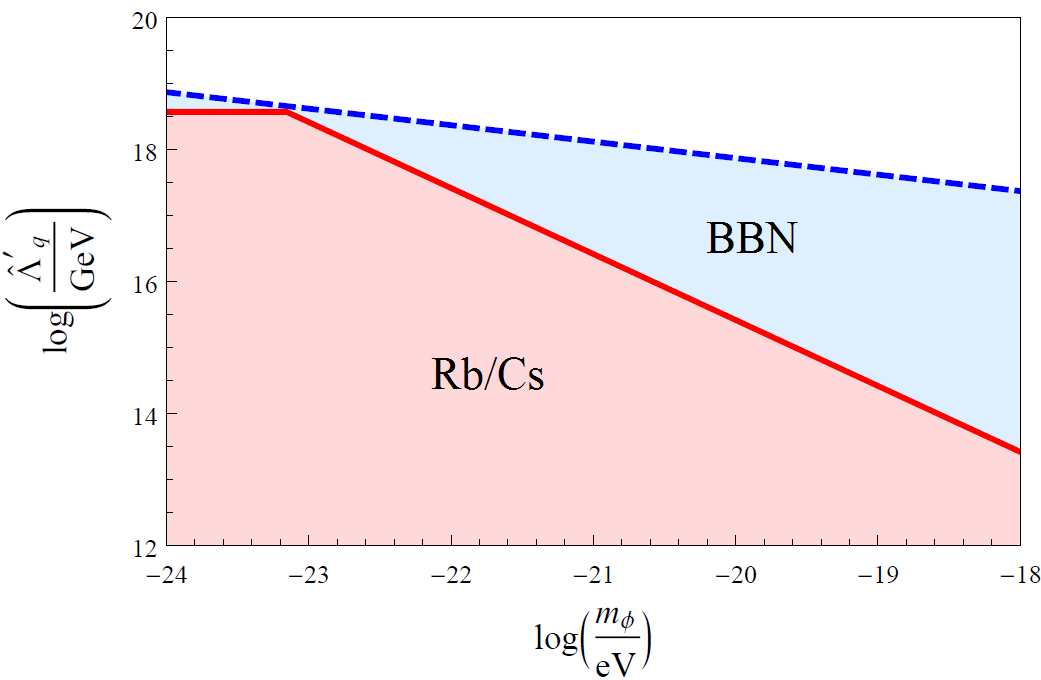} 
\caption{(Color online) From top to bottom:~Limits on the linear interaction of $\phi$ with the Higgs boson (assuming $b=0.2$), and on the quadratic interactions of $\phi$ with the photon and light quarks, as functions of the dark matter particle mass $m_\phi$. 
The region in red corresponds to constraints derived in the present work using recent Rb/Cs atomic spectroscopy data of \cite{Bize2016}. 
The region in yellow corresponds to constraints derived from Dy atomic spectroscopy data of \cite{Budker2015} (the constraints on the Higgs interaction parameter are derived in the present work, while the constraints on the photon interaction parameter were derived in \cite{Stadnik2015DM-VFCs}). 
The region in grey corresponds to constraints from fifth-force searches \cite{Pospelov2010}. 
The region in blue corresponds to constraints from consideration of the primordial $^{4}$He abundance produced during Big Bang nucleosynthesis \cite{Stadnik2015DM-VFCs}. 
The region in purple corresponds to constraints from consideration of cosmic microwave background angular power spectrum measurements \cite{Stadnik2015DM-VFCs}. 
The quark interaction parameters that appear in the bottom graph are defined in-text.
} 
\label{fig:Higgs_linear_space} 
\end{center}
\end{figure}

From the Dy data in \cite{Budker2015} and the Rb/Cs data in \cite{Bize2016}, we obtain the limits on the interaction parameter $A$, as shown in Fig.~\ref{fig:Higgs_linear_space}, assuming that $b=0.2$. 
For Dy, the most sensitive limit is $A \lesssim 1.4 \times 10^{-10}$ eV for $m_\phi \approx 6 \times 10^{-23}$ eV, while for Rb/Cs, the most sensitive limit is $A \lesssim 4 \times 10^{-13}$ eV for $m_\phi \approx 1.4 \times 10^{-23}$ eV. 
The Rb/Cs bound on $A$ improves on the existing limits from fifth-force searches \cite{Pospelov2010}, which are $A \lesssim 10^{-10}$ eV for the mass range of interest, by up to $2-3$ orders of magnitude.

Also from the Rb/Cs data in \cite{Bize2016}, we obtain the limits on the interaction parameters $\Lambda'_\gamma$ and $\hat{\Lambda}'_q$, as shown in Fig.~\ref{fig:Higgs_linear_space}, where $\hat{\Lambda}'_q$ is defined as $\left( \hat{\Lambda}'_q \right)^2 = \left[\left( \Lambda'_u \right)^2 \left( \Lambda'_d \right)^2 (m_d + m_u)\right] / \left[ m_d \left( \Lambda'_u \right)^2 + m_u \left( \Lambda'_d \right)^2  \right] $. 
The most sensitive limits are $\Lambda'_\gamma \gtrsim 2 \times 10^{19}$ GeV and $\hat{\Lambda}'_q \gtrsim 4 \times 10^{18}$ GeV for $m_\phi \lesssim 7 \times 10^{-24}$ eV. 
The Rb/Cs bound on $\Lambda'_\gamma$ improves on existing limits from Dy spectroscopy \cite{Stadnik2015DM-VFCs}, which are $\Lambda'_\gamma \gtrsim 3 \times 10^{18}$ GeV for $m_\phi \lesssim 3 \times 10^{-23}$ eV, and from Big Bang nucleosynthesis (BBN) \cite{Stadnik2015DM-VFCs}, which are $\Lambda'_\gamma \gtrsim \left(4 \times 10^9~\textrm{eV}^2 / m_\phi \right) \left[m_\phi / (3 \times 10^{-16}~\textrm{eV}) \right]^{3/4} $ for the mass range of interest, by up to a factor of 6. 
The Rb/Cs bound on $\hat{\Lambda}'_q$ in the most sensitive region is comparable to existing limits from BBN on the parameter $\tilde{\Lambda}'_q$, defined as $\left( \tilde{\Lambda}'_q \right)^2 = \left[\left( \Lambda'_u \right)^2 \left( \Lambda'_d \right)^2 (m_d - m_u)\right] / \left[ m_d \left( \Lambda'_u \right)^2 - m_u \left( \Lambda'_d \right)^2 \right] $, which are $\tilde{\Lambda}'_q \gtrsim \left(2 \times 10^{10}~\textrm{eV}^2 / m_\phi \right) \left[m_\phi / (3 \times 10^{-16}~\textrm{eV}) \right]^{3/4} $ for the mass range of interest.

\textbf{Conclusions.} ---
We have calculated the effects of possible interactions of low-mass (sub-eV) spin-0 dark matter particles, which form a coherently oscillating classical field $\phi = \phi_0 \cos(m_\phi t)$, with the Standard Model sector, including the linear interaction of $\phi$ with the Higgs boson, on atomic and molecular transitions. 
Using recent atomic clock spectroscopy measurements, we have derived new limits on the linear interaction of $\phi$ with the Higgs boson, as well as its quadratic interactions with the photon and light quarks. 
For the linear interaction of $\phi$ with the Higgs boson, the new limits from Rb/Cs improve on existing constraints by up to $2-3$ orders of magnitude. 
For the quadratic interaction of $\phi$ with the photon, the new limits from Rb/Cs improve on existing constraints by up to a factor of 6, while for the quadratic interaction of $\phi$ with the light quarks, the new limits from Rb/Cs in the most sensitive region are comparable to existing constraints. 

Further improvements in sensitivity to dark matter interaction parameters may come from an analysis of other existing atomic clock spectroscopy data, which include the systems Al$^{+}$/Hg$^{+}$ \cite{Rosenband2008clock}, Yb$^{+}$(E3)/Cs \cite{Peik2014clock}, Yb$^{+}$(E3)/Yb$^{+}$(E2) \cite{Gill2014clock} and Sr/Yb/Hg \cite{Katori2015clock}. 
Finally, regarding searching for the possible linear interaction of $\phi$ with the Higgs boson, we note that an atomic optical/hyperfine frequency ratio comparison would offer a particularly sensitive platform, while certain molecular transitions may offer even higher sensitivity.

\textbf{Acknowledgements.} --- 
This work was supported by the Australian Research Council.



\begin{thebibliography}{99}
\bibitem{Footnote1} In order for spin-0 DM to form a classical field, there must be a large number of such DM particles within the reduced de Broglie volume, i.e., $n_\phi \left( \lambda_{\textrm{dB}}/2\pi \right)^3 \gg 1$. For spin-0 DM particles which saturate the observed local cold DM content, this requirement translates into the upper mass bound:~$m_\phi \lesssim 0.1$ eV.
\bibitem{Footnote3} For the super-renormalisable interaction (\ref{lin_couplings_phi_Higgs}), the lowest-order radiative corrections give a logarithmically-divergent correction to $m_\phi^2$, in contrast to the usual quadratically-divergent corrections. 
This allows one to avoid problems with technical naturalness for $A \lesssim m_\phi$. 
If $(A / m_\phi)^2 < 2\lambda$, where $\lambda = m_h^2/2\left<h\right>^2 \approx 1/8$, then the minimum of the Higgs potential is stable; 
otherwise, additional non-linear $\phi^4$ terms must be introduced.
\bibitem{Footnote4} In Eqs.~(\ref{clock_frequency_ratio_lin}) and (\ref{clock_frequency_ratio_quad}), we have chosen a parametrisation in terms of fundamental quark interaction parameters, rather than a parametrisation in terms of effective nucleon parameters.
\end{thebibliography}


\end{document}